# Background-free coherent anti-stokes Raman spectroscopy and microscopy by dual-soliton pulse generation

Kun Chen*, Tao Wu, Haoyun Wei, Yan Li

*Key Lab of Precision Measurement Technology & Instrument, Department of Precision Instrument, Tsinghua University, Beijing 100084, China*
*Corresponding author: chen-k13@mails.tsinghua.edu.cn

We propose an all-fiber-generated dual-soliton pulses based scheme for the background-free detection of coherent anti-Stokes Raman spectroscopy under the spectral focusing mechanism. Due to the strong birefringence and high nonlinearity of a polarization-maintaining photonic crystal fiber (PM-PCF), two redshifted soliton pulses can be simultaneously generated relying on high-order dispersion and nonlinear effects along two eigenpolarization axes in the anomalous dispersion regime, while allowing feasible tunability of the frequency distance and temporal interval between them. This proposed scheme, termed as DS-CARS, exploits a unique combination of slight frequency-shift and advisable temporal walk-off of this two soliton pulses to achieve robust and efficient suppression of nonresonant background with compact all-fiber coherent excitation source. Capability of the DS-CARS is experimentally demonstrated by the background-free CARS spectroscopy and unambiguous CARS microscopy of polymer beads in the fingerprint region.

In the last decade, coherent anti-Stokes Raman scattering (CARS) microscopy has being used increasingly as a unique microscopic tool in biophysics, biology and material sciences [1-3]. However, a background from nonresonant (NR) CARS contributions, carrying no chemically specific information severely limits its sensitivity and specificity. The NR background can further distort and even overwhelm the resonant signal of interest, especially in the fingerprint spectral region where molecules present their unique vibrational signature with lower cross section than the frequently used C–H stretch [2]. Several methods have been explored to overcome this problem including phase-and polarization control CARS [4], time-resolved CARS [5], Fourier transform CARS [6], heterodyne interferometric CARS [7] and frequency modulation CARS (FM-CARS) [8, 9]. In particular FM-CARS technique has demonstrated its efficient suppression of NR CARS background. However, the reported implementations require hardware modification to provide a second pump laser [8] or modulation of the time delay in the spectral focusing scheme, yielding a complex and costly setup [9]. In addition to the NR background, the complexity of the excitation sources also prevents a broader use of CARS microscopy. The current gold-standard laser system for CARS microscopy is synchronized picosecond (ps) mode-locked solid-state oscillators [10], or synchronously pumped ps OPOs [11]. Free space lasers generally require a stable environment or active feedback, whereas an all-fiber source could be used in less favorable environments while being more compact and inexpensive. Recently, Er- and Yb-doped fiber lasers combined with highly nonlinear fiber or PCF have been put to use in a new type of all-fiber-optic coherent Raman microscopy [12]. Meanwhile, transform-limited (TL) output femtosecond pulses can also be achieved by taking advantage of the nonlinearities of a PCF in the anomalous dispersion regime due to the balance between anomalous dispersion and the Kerr nonlinearity [13]. It opens the door to spectroscopic applications and hyperspectral imaging, offering a feasible route at solving the precompensation problem while enabling tunability [14]. However, none of studies focus on highly birefringent PCFs and the current fiber-based CARS techniques still suffers from the NR background issue. The purpose of this Letter is to employ a new dual-soliton excitation scheme to perform intrinsically efficient background-free detection of CARS spectroscopy and microscopy based on all-fiber excitation source. We call the scheme dual-soliton CARS detection (DS-CARS). It exploits two temporal walk-off and slight frequency-shift solitons as Stokes pulses simultaneously generated in a highly birefringent and nonlinear PM-PCF. We will first detail the nature of the dual-soliton generation, and then we will introduce the principle of the DS-CARS scheme. We will finally validate the DS-CARS scheme on polymer samples by background-free CARS spectroscopy and unambiguous CARS imaging.

The basic mechanism underlying the generation of supercontinuum in PCF fibers is the fission of higher-order solitons due to high-order dispersion and nonlinear effects (four-wave mixing, self-phase modulation and cross-phase modulation). With respect to PM-PCF, the strong ellipticity of the core is necessary for the preservation of the state of polarization of the pulses, also inducing the high birefringence of the fiber [15]. On this basis, the birefringence has the potential to simultaneously generate two continua with orthogonal polarizations, allowing for an extra degree of freedom in tuning the properties of the supercontinuum in the application of CARS microscopy. To qualitatively analysis the evolution of both frequency components and temporal pulse along the fast and slow axes of strongly birefringent fiber, numerical simulation is firstly conducted, governed by the Ginzburg-Landau equations without the gain terms [16]

$$\frac{\partial u}{\partial z} = i\beta u - \delta \frac{\partial u}{\partial t} - \frac{ik''}{2}\frac{\partial^2 u}{\partial t^2} + \frac{ik'''}{6}\frac{\partial^3 u}{\partial t^3} + i\gamma(|u|^2 + \frac{2}{3}|v|^2)u + \frac{i\gamma}{3}v^2 u^* \quad (1)$$

$$\frac{\partial v}{\partial z} = -i\beta v + \delta \frac{\partial v}{\partial t} - \frac{ik''}{2}\frac{\partial^2 v}{\partial t^2} + \frac{ik'''}{6}\frac{\partial^3 v}{\partial t^3} + i\gamma(|v|^2 + \frac{2}{3}|u|^2)v + \frac{i\gamma}{3}u^2 v^* \quad (2)$$

where $u$ and $v$ are the normalized envelopes of the optical pulses in the two orthogonal polarized modes. $2\beta = 2\pi\Delta n/\lambda$ is the wave-number difference between the two modes. $2\delta = 2\beta\lambda/2\pi c$ is the inverse group velocity difference. $k''$ is the second-order dispersion coefficient; $k'''$ is the third-order dispersion coefficient and $\gamma$ represents the nonlinear coefficient. All of the parameters used in our simulations possibly match the experimental conditions. Specifically, a commercial PM-PCF (NKT Photonics SC-5.0–1040-PM) acts as the high birefringent fiber in practice. So, $k''$=-4.2 ps$^{\wedge}2$/km, $k'''$=0.0681 ps$^{\wedge}3$/km, $\gamma$=10 W$^{-1}$km$^{-1}$, beat length=6.3 mm and PCF length=193 cm. The input pulse is slightly chirped (~ 178fs) with peak power ~2.5 kW from a home-built Yb-doped fiber laser (100MHz @1060nm), which will be detailed in the next section.

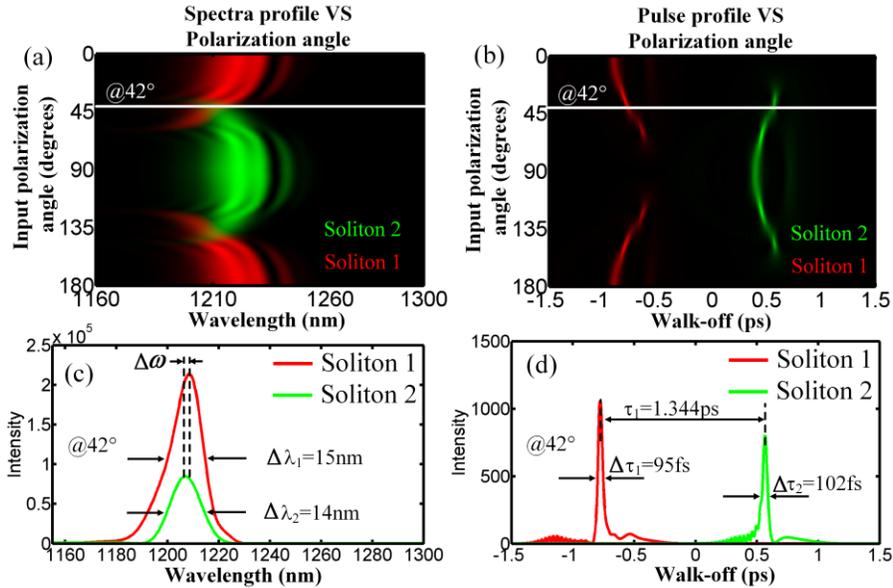

Fig. 1. Simulated results of the spectra and pulse profiles at the output of the PM-PCF after filtering with a longpass filter (>1150nm). The 0° position was aligned with the fast axis of the fiber. (a) and (b) are the evolution of the generated spectra and pulse as the input polarization is rotated. Red part: soliton 1; green part: soliton 2. (c) and (d) are the spectra and pulse profiles when the input polarization is 42° as indicated with white lines in (a) and (b). Difference of center frequency between soliton 1 and 2 is about $\Delta\omega$=1.5nm.

By rotating the input polarization while maintaining the injected power, numerically simulated results in Fig. 1 qualitatively display the evolution of the generated spectra and soliton pulse profiles along the fast and slow axis of the birefringent PCF, respectively. When the input polarization is oriented along one of the polarization axes (0° or 90°), spectral components only generate in one orientation in Fig. 1(a), which corresponds to only one soliton pulse in Fig. 1(b). When the input polarization isn't parallel to the fast or slow axes, contour plots of the spectrum in (a) further provides an unambiguous proof that the generated spectrum is a linear combination of two continua generated separately along the two principal axes. Indeed, spectral components along one axis are not coupled to those in the other axis. This is attributable to the different dispersion characteristics of the two eigenpolarizations, thus leading to the temporal walk-off of this dual-soliton in (b). The most notable feature of the dual-soliton behavior is the intrinsic but tunable frequency distance and temporal interval between them, which are the key points to implement background free CARS spectroscopy and microscopy in the DS-CARS scheme. This temporal interval can be naturally applied to a time-sweep chirped CARS scheme, like the spectral focusing technique [17], which establishes an ingenious time-frequency transform to allow the subsequent shifted excitation relying on an appropriate frequency distance, by scanning the time delay between the pump and the dual-soliton pulses. Shifted excitation with two excitation wavelengths shifted by an amount comparable to the bandwidth of the measured Raman bands (typically 10 cm$^{-1}$ in the fingerprint region) has been demonstrated as an effective way to eliminate the large backgrounds in the FM-CARS method [9]. As an example, Fig. 1(c) and (d) presents a typical dual-soliton output by carefully adjusting the input polarization angle at 42°. Soliton 2 shifts just about 1.5nm (~10 cm$^{-1}$) relative to soliton 1, towards to shorter wavelength. Besides, the temporal interval is about 1.344ps. The dual-soliton therefore offers a unique combination of slight frequency-shift and advisable walk-off, making it a versatile source for background-free CARS under spectral focusing framework. Besides, the bandwidth of soliton 1 and 2 are 15nm and 14 nm corresponding to the pulse durations 95fs and 102fs, respectively. Considering the pulse profile is hyperbolic-secant squared when they leave the PCF, both soliton 1 and 2 are nearly transform-limited, benefiting the chirp adjustment of the pump and Stokes beams to be almost equal.

In the spectral focusing CARS mechanism, a single Raman level can be excited by multiple pairs of pump and Stokes frequency components, accompanying with a large NR nonspecific four-wave-mixing signal at the same frequency as schematically illustrated in Fig. 2(a) and (b). The total CARS response can be described as the sum of resonant third-order nonlinear polarization $P_R^{(3)}(\omega_{as})$ and the NR one $P_{NR}^{(3)}(\omega_{as})$. The detected CARS signal is therefore proportional to [2]

$$I_{CARS}(\omega_{as}) \propto \left| P_{NR}^{(3)}(\omega_{as}) + P_R^{(3)}(\omega_{as}) \right|^2 \qquad (3)$$
$$= \left| P_{NR}^{(3)}(\omega_{as}) \right|^2 + \left| P_R^{(3)}(\omega_{as}) \right|^2 + 2 P_{NR}^{(3)}(\omega_{as}) \text{Re}\{P_R^{(3)}(\omega_{as})\}$$

$P_{NR}^{(3)}(\omega_{as})$ is real, chemically-nonspecific typically assumed to be slowly varying in frequency (although not necessarily constant). Note that the third term is the CARS signal multiplied by the NR background, which is referred to as heterodyne amplification. In the DS-CARS scheme, the Raman level $\Omega$ can be excited twice while scanning the interpulse delay between the pump and the Stokes beam shown in Fig. 2(a). So, we can acquire CARS signals at different time-delay: $I_{CARS1}(\omega_{as}, \omega_{as1})$ and $I_{CARS2}(\omega_{as}, \omega_{as2})$. In general, the maximum material response will occur at

$$\omega_{as1} = \omega_{pr1} + \omega_{p1} - \omega_{s1} = \omega_{pr1} + \Omega$$
$$\omega_{as2} = \omega_{pr2} + \omega_{p2} - \omega_{s2} = \omega_{pr2} + \Omega \qquad (4)$$

From Eq. (4), it can be concluded that the emitted anti-Stokes frequency will basically track with the probe frequency after the built-up of the Raman coherence by the first-come pump and Stokes pulses. Due to the intrinsic frequency shift of soliton 2 (Fig. 2(a)), we have

$$\omega_{as2} - \omega_{as1} = \omega_{pr2} - \omega_{pr1} = \Delta\omega \qquad (5)$$

This frequency difference allow us to exhibit a shifted Raman difference spectrum between $I_{CARS1}(\omega_{as}, \omega_{as1})$ and $I_{CARS2}(\omega_{as}, \omega_{as2})$ under the assumption of $P_R^{(3)}(\omega_{as}) \ll P_{NR}^{(3)}(\omega_{as})$ in the fingerprint regime [9]

$$\Delta I_{CARS}(\omega_{as}) = I_{CARS1}(\omega_{as}, \omega_{as1}) - I_{CARS2}(\omega_{as}, \omega_{as2})$$
$$\propto 2 P_{NR}^{(3)}(\omega_{as})(\text{Re}\{P_R^{(3)}(\omega_{as}, \omega_{as1})\} - \text{Re}\{P_R^{(3)}(\omega_{as}, \omega_{as2})\}) \qquad (6)$$
$$= 2 P_{NR}^{(3)}(\omega_{as}) \text{Re}\{P_R^{(3)}(-\omega_{as})\} \otimes [\delta(\omega_{as} - \omega_{as1}) - \delta(\omega_{as} - \omega_{as2})]$$

Thus, the difference spectrum is simply the real part CARS spectrum convolved with the difference of the two delta functions excluding the underlying broad NR background. Applying the "Difference Deconvolution" method [18] to Eq. (6), we obtain

$$\text{Re}\{P_R^{(3)}(-\omega_{as})\}$$
$$\propto F^{-1}[F(\Delta I_{CARS}(\omega_{as})) / F(\delta(\omega_{as} - \omega_{as1}) - \delta(\omega_{as} - \omega_{as2}))] \qquad (7)$$

where $F$ denotes forward Fourier transform, and $F^{-1}$ denotes the inverse. The desired Raman signal is finally derived from the connection between the squared-module and the real part of the third-order nonlinear polarization $P_R^{(3)}(\omega_{as})$

$$k \times \left| P_R^{(3)}(\omega_{as}) \right|^2 \times \text{Re}\{P_R^{(3)}(\omega_{as})\} = \frac{d(\left| P_R^{(3)}(\omega_{as}) \right|^2)}{d\omega_{as}} \qquad (8)$$

$$\left|P_R^{(3)}(\omega_{as})\right|^2 \propto \exp[k \times \int \text{Re}\{P_R^{(3)}(\omega_{as})\}d\omega_{as}] \quad (9)$$

where, $k$ is a constant coefficient. It should be noted that we attain the squared-module of the resonant third-order nonlinear susceptibility $\chi_R^{(3)}$, which is proportional to spontaneous Raman response (spectra). Background-free CARS signals, without any distortions and dissipative characteristics, can be demonstrated theoretically from Eq. (9) under the Dual-soliton pulses framework.

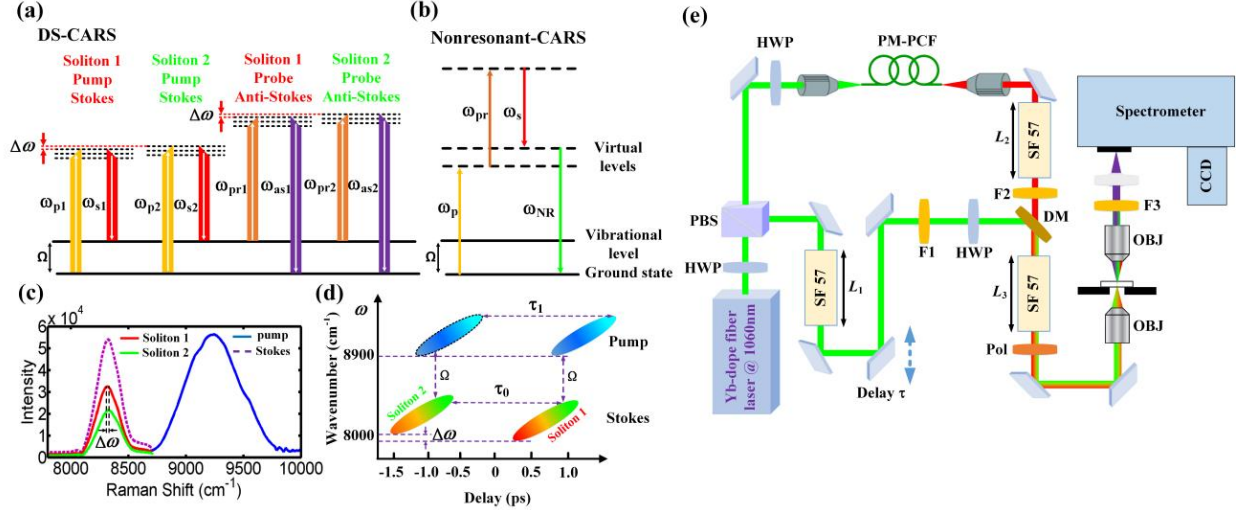

Fig. 2. (a) Energy-level scheme of the DS-CARS process under spectral focusing mechanism. (b) Energy level diagrams of nonresonant CARS. (c) The output spectra of Stokes and pump; Stokes profile is a linear combination of soliton 1 and 2. (d) Principle of the spectral focusing DS-CARS scheme. (e) Sketch of the experimental setup: HWP, half wave-plate; PBS, polarizing beam splitter; PCF, photonic crystal fiber; F1, F2, long-pass filter; F3, short-pass filter; SF57, SF-57 glass rod; DM, dichroic mirror; OBJ, objective lenses; $L_1$=30mm, $L_2$=108mm, $L_3$=200mm.

In practice, the implementation of DS-CARS proceeds as presented in Fig. 2(e). As mentioned above, a home-built Yb-doped fiber laser with an output of 2W average power at 100 MHz repetition rate was used as the primary source. This laser output was split into the pump and the Stokes paths. The continuum redshifts to about 1200nm so as to probe the Raman shift of polystyrene around 1001 cm$^{-1}$, by adjusting the input power to the PCF using a half wave-plate and polarizing beam splitter combination. By rotating the input polarization, the generated spectrum at the fiber output actually consists of a superposition of the spectra generated independently by the two eigenpolarization modes, which is termed as Soliton 1 (red line) and Soliton 2 (green line) in Fig. 2(c). These results are in good agreement with the simulated contour plots in Fig. 1(a) and (c). The temporal interval of the two walk-off TL soliton pulses is about $\tau_0$=1.43ps as illustrated in Fig. 2(d), which can be characterized by autocorrelator. Meanwhile, the spectral components generated in either axis are not coupled to those in the other axis, leading to the tunable difference of center frequency between two soliton pulses ($\Delta\omega$=10 cm$^{-1}$, illustrated in Fig. 2(c)). Different length of SF57 ($L_1$, $L_2$ and $L_3$) are added to ensure that the same amount of linear positive chirp are applied to both pump and Stokes beam. After traveling through these high-index glass rod, the pulse durations of pump and Stokes pulses are FWHM =2.2ps and FWHM =1.4ps, respectively. A motorized delay line is inserted into the pump beam to scan the delay between the two pulse trains. These two beams are overlapped in space by a dichroic mirror and sent into a custom-made scanning microscope with a near-infrared optimized excitation objective lens (NA=0.65, LCPLN50XIR, Olympus) and a collection one (NA=0.65, LCPLN50XIR, Olympus). The signal at anti-Stokes frequency is detected in the forward direction by means of an imaging spectrograph (IsoPlane160, Princeton Instruments) attached with a back-illuminated, deep depletion CCD (PIXIS1024BR, Princeton Instruments). For imaging, the sample is raster scanned.

As a first experimental demonstration, we performed DS-CARS spectroscopy. Polystyrene (PS) beads of 20 $\mu m$ diameter mixed with 98% water are firstly prepared. Agarose (Biowest) is then added to the solution to a final concentration of 30% by weight. A drop of solution was pipetted inside a 120 $\mu m$ thick imaging spacer (Grace$^{TM}$Bio-Lab SecureSeal$^{TM}$) glued on a glass coverslip (1 mm thick) in order to create a chamber, which was sealed by a second coverslip (0.13 mm thick) on top. Glass coverslip as well as high concentrations of water and agarose presents NR signal only, so we can evaluate the DS-CARS modality in a more challenging and realistic case. Fig. 3(a) shows the measured CARS spectrograms and (b) is the measured NR signals along the dashed white line in (a) indicating the delay = 0. We are bringing clear evidences that it still presents a significant NR signals off resonance (delay=0), which is a major impediment to the resonant signals. The CARS spectrum is extracted from the three-dimensional spectrograms as the amplitude along the time-delay by integrating the total power of each anti-Stokes spectrum from the spectrometer. As shown in Fig. 3(c), CARS signals produced from both soliton 1 and soliton 2 present strong NR backgrounds, mostly coming from glass coverslip, agarose and water, further leading to distortion and misunderstanding in the CARS microscopy. In some cases, like weak Raman scatterers in fingerprint spectral region the NR signal would overwhelm the resonant signal. Besides, for twice excitation of the same polystyrene Raman levels, the pump pulse should scan the two Stokes soliton pulses at different delay-time and the delay interval is about $\tau_1$=1.47ps. It is worth noting that the delay interval $\tau_1$ is not equal to the temporal interval of two TL soliton pulses $\tau_0$, due to the slight difference of spectral components between soliton 1 and soliton 2: $\Delta\omega$ in Fig. 2(d). This slight time-domain shift ($\tau_1$-$\tau_0$) can convert into the frequency-domain Raman peak shift through the time-frequency transform, under the spectral focusing mechanism. On this basis the underlying broad NR background can be subtracted from the difference between these two CARS signals, and the spectral shape of its output looks close to the shape of spontaneous Raman scattering (Fig. 3(e)). Even the difference CARS is more resolvable and almost background–free, subsequent Raman-retrieve method using Eq. (9) can further extract the net resonant nonlinear susceptibility, providing more faithful and baseline-corrected CARS spectra. For 20 $\mu$m PS beads (Fig. 3(f)), the acquired DS-CARS spectrum has a signal-to-noise ratio up to 100. Spectral position and FWHM of the two close Raman resonance at 1001 cm$^{-1}$ and 1032 cm$^{-1}$ match very well with the spontaneous spectrum, allowing clear assignment of the resolved Raman lines to the corresponding vibrational transitions.

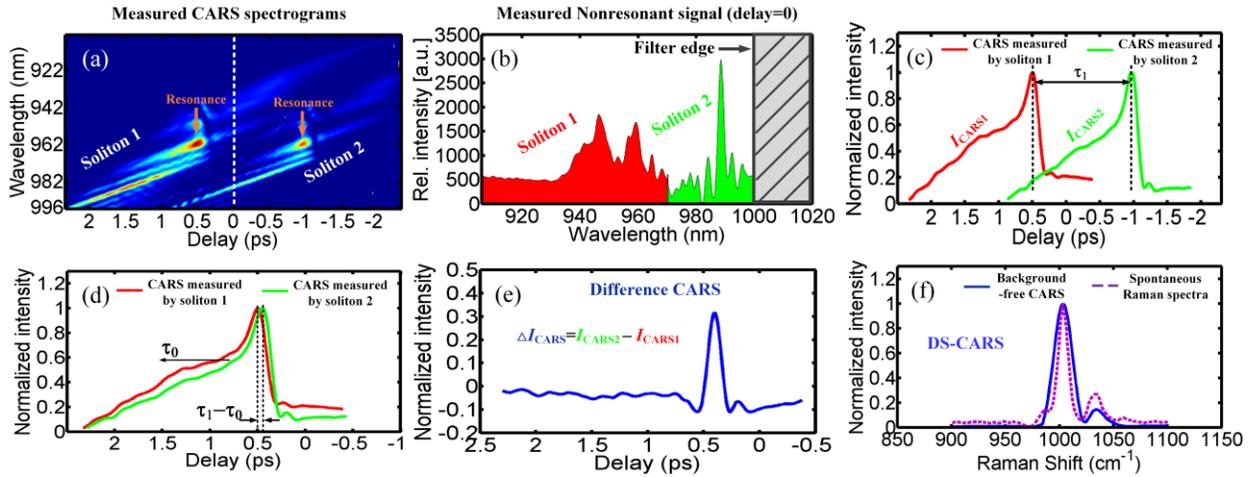

Fig. 3. (a) Measured CARS as a function of pulse delay and the signal wavelength. (b) Measured NR signals along the dashed white line in (a) indicating the delay = 0. (c) CARS spectra generated from soliton 1 and soliton 2. (d) Shift $I_{CARS2}$ towards $I_{CARS1}$ by the amount of $\tau_0$. (e) Difference CARS signals between $I_{CARS2}$ and $I_{CARS1}$ in time domain. (f) Retrieved and baseline-corrected background-free DS-CARS.

We wish now to evaluate the DS-CARS modality in label-free microscopy case where the CARS imaging is most advantageous compared with the conventional fluorescence microscopy. Fig. 4(a) shows normal forward CARS images using only soliton 1 or soliton 2 as Stokes pulse and the background-free CARS images using DS-CARS scheme. All of the images depend on the intrinsic contrast of 20 $\mu$m polystyrene beads at the strongly resonant Raman level ($\Omega$=1001 cm$^{-1}$) and the weakly resonant one ($\Omega$=1032 cm$^{-1}$). The normal CARS images (using soliton 1 or soliton 2 alone) show a resonant signal only slightly above the NR background, especially when it comes to the weak Raman resonance at 1032 cm$^{-1}$. Cross-sectional profiles of the images along the indicated white line in Fig. 4(c) clearly present that NR background almost overwhelm the resonant signal at 1032 cm$^{-1}$ and the image contrast is as low as 0.07. Besides, distortions and fake peaks can mislead the interpretation of local chemical components at different spatial locations (see the larger boxed regions). However, the DS-CARS images both at strong and weak resonance are completely background-free and make the beads stand out more clearly in Fig. 4(a). The corresponding cross-sectional profiles of the images demonstrate considerably better contrast ((b) and (c)), illustrating the achieved contrast enhancement of a factor of ~2 and 8 compared to the normal spectral focusing CARS at 1001 cm$^{-1}$ and 1032 cm$^{-1}$, respectively. This contrast enhancement also demonstrates that DS-CARS is most advantageous when the resonant signal is comparable with or smaller than the NR background at 1032 cm$^{-1}$. Most biomedical CARS imaging has been performed at the symmetric CH$_2$ stretching bands, which is uniquely isolated from other peaks and very strong in lipid samples due to its abundance in lipid acyl chains. Since the CARS cross section is quadratically proportional to that of spontaneous Raman scattering, it implies that the CH$_2$ stretch is significantly stronger than most peaks in the fingerprint region. However, the fingerprint region is an important frequency range and offers rich chemical information [2]. So, the DS-CARS is most applicable to the vibrational fingerprint region of weak Raman scatterers.

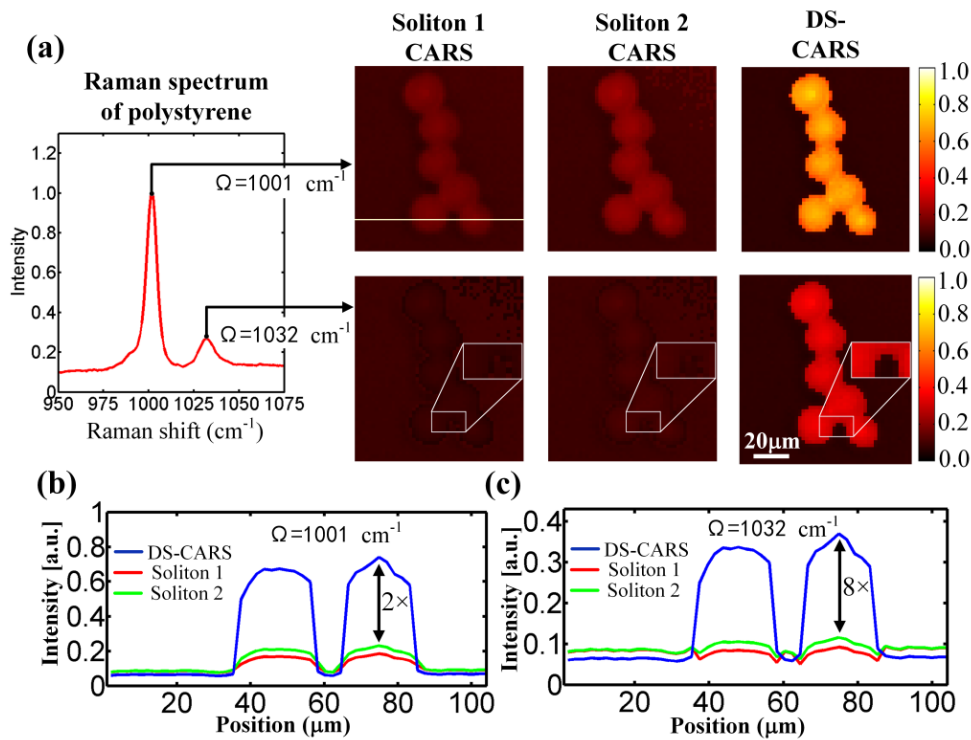

Fig. 4. (a) Measured Soliton 1 CARS, Soliton 2 CARS and DS-CARS images of 20 $\mu$m polystyrene beads immersed in a gel of agarose and water for two Raman bands at $\Omega$=1001 cm$^{-1}$ and 1032 cm$^{-1}$, respectively. Image 55×50 pixels, $I_{pump}$=30 mW, $I_{Soliton\ 1}$=1.8 mW, $I_{Soliton\ 2}$=1 mW. (b) and (c) Corresponding cross-sectional profiles of the images along the indicated white line show the magnitude of NR background suppression, illustrating the achieved contrast enhancement of a factor of ~2 and 8 compared to the normal spectral focusing with one soliton pulse at $\Omega$=1001 cm$^{-1}$ and 1032 cm$^{-1}$, respectively.

We have reported on a method to perform background-free CARS spectroscopy and microscopy taking advantage of the intrinsically temporal walk-off and slight frequency-shift dual-soliton pulses generated in a high birefringent PM-PCF under the spectral focusing mechanism. Not only does it provide true background-free CARS imaging and spectroscopy, it also leads to several fold increase in the image contrast. The experimental implementation of DS-CARS is straightforward and takes full advantage of the all-fiber-based source technology which could be used in less favorable environments while being more compact and inexpensive. All these features are unique to the DS-CARS scheme and differ from previously reported efforts aimed at suppressing the NR background signals in CARS. Whereas the background-free and contrast enhancement has been demonstrated here at ring breathing (1001 cm$^{-1}$) and C–O stretch (1032 cm$^{-1}$), it would also work in other frequently used vibrational Raman bands offering chemical contrast or specificity like C = C stretch (1665 cm$^{-1}$), CH$_2$ stretch (2884 cm$^{-1}$) and O–H stretch (3228 cm$^{-1}$). Because our all-fiber based source system with supercontinuum generation can intrinsically offer the feasible tunability of excitation for CARS. We believe DS-CARS can bring valuable sensitivity and specificity enhancement in label-free molecular imaging and spectroscopy.

### Acknowledgements
This work is supported by the State Key Laboratory of Precision Measurement Technology & Instrument of Tsinghua University and the Tsinghua University Initiative Scientific Research Program. Special thanks and appreciation go to Pan Wang for help with the simulation of dual-soliton generation. We also thank Lukas Bruckner (Physikalisch-Chemisches Institut, Germany) and Tao Chen (Biodynamic Optical Imaging Center, Peking University, China) for the very helpful discussions about the spectral focusing technique.

### Reference
[1] A. Zumbusch, G. R. Holtom, and X. S. Xie, Phys. Rev. Lett. **82**, 4142 (1999).

[2] C. H. Camp, Y. J. Lee, J. M. Heddleston, C. M. Hartshorn, A. R. H. Walker, J. N. Rich, J. D. Lathia, and M. T. Cicerone, Nat. Photonics **8**, 627 (2014).
[3] M. O. Scully, G. W. Kattawar, R. P. Lucht, T. Opatrny´, H. Pilloff, A. Rebane, A. V. Sokolov, and M. S. Zubairy, Proc. Natl. Acad. Sci. U.S.A. **99**, 10994 (2002).
[4] D. Oron, N. Dudovich, and Y. Silberberg, Phys. Rev. Lett. **90**, 213902 (2003).
[5] R. Selm, M. Winterhalder, A. Zumbusch, G. Krauss, T. Hanke, A. Sell, and A. Leitenstorfer, Opt. Lett. **35**, 3282 (2010).
[6] M. Cui, M. Joffre, J. Skodack, and J. P. Ogilvie, Opt. Express **14**, 8448 (2006).
[7] D. L. Marks, C. Vinegoni, J.S. Bredfeldt, and S.A. Bopparta, Appl. Phys. Lett. **85**, 5787 (2004).
[8] F. Ganikhanov, C. L. Evans, B. G. Saar, and X. S. Xie, Opt. Lett. **31**, 1872 (2006).
[9] B. Chen, J. Sung, and S. Lim, J. Phys. Chem. B **114**, 16871 (2010).
[10] F. Ganikhanov, S. Carrasco, X. S. Xie, M. Katz, W. Seitz, and D. Kopf, Opt. Lett. **31**, 1292 (2006).
[11] O. Burkacky, A. Zumbusch, C. Brackmann, and A. Enejder, Opt. Lett. **31**, 3656 (2006).
[12] C. W. Freudiger, W. Yang, G. R. Holtom, N. Peyghambarian, X. S. Xie, and K. Q. Kieu, Nat. Photonics **8**, 153 (2014)
[13] E. Andresen, P.l Berto, and H. Rigneault, Opt. Lett. **26**, 2387 (2011)
[14] E. Andresen, V. Birkedal, J. Thøgersen, and S. R. Keiding, Opt. Lett. **31**, 1328 (2006).
[15] M. Lehtonen, G. Genty, and H. Ludvigsen, Appl. Phys. Lett. **82,** 2197 (2003).
[16] D. Y. Tang, H. Zhang, L. M. Zhao, and X. Wu, Phys. Rev. Lett. **101**, 153904 (2008).
[17] T. Hellerer, A. M. K. Enejder, and A. Zumbuscha, Appl. Phys. Lett. **85**, 25 (2004).
[18] J. Zhao, M. M. Carrabba, and F. Allen, Appl. Spec. **56**, 834 (2002).